\documentclass[a4paper,11pt]{article}
\usepackage{pos}

\usepackage{siunitx}
\usepackage{amsmath}
\usepackage{lineno}

\title{Combining Maximum-Likelihood with Deep Learning for Event Reconstruction in IceCube}

\ShortTitle{Combining Maximum-Likelihood with Deep Learning}

\author{The IceCube Collaboration \\{\normalsize \normalfont(a complete list of authors can be found at the end of the proceedings)}}

\emailAdd{mirco.huennefeld@tu-dortmund.de}

\abstract{

The field of deep learning has become increasingly important for particle physics experiments, yielding a multitude of advances, predominantly in event classification and reconstruction tasks. Many of these applications have been adopted from other domains. However, data in the field of physics are unique in the context of machine learning, insofar as their generation process and the laws and symmetries they abide by are usually well understood. Most commonly used deep learning architectures fail at utilizing this available information. In contrast, more traditional likelihood-based methods are capable of exploiting domain knowledge, but they are often limited by computational complexity.
In this contribution, a hybrid approach is presented that utilizes generative neural networks to approximate the likelihood, which may then be used in a traditional maximum-likelihood setting. Domain knowledge, such as invariances and detector characteristics, can easily be incorporated in this approach. The hybrid approach is illustrated by the example of event reconstruction in IceCube.
\\

\vspace{4mm}
{\bfseries Corresponding authors:}
Mirco H\"{u}nnefeld$^{1*}$\\
{$^{1}$ \itshape TU Dortmund University}\\
$^*$ Presenter

\FullConference{37$^{\rm{th}}$ International Cosmic Ray Conference (ICRC 2021)\\
		July 12th -- 23rd, 2021\\
		Online -- Berlin, Germany}

}

\begin{document}
\maketitle


\section{The Importance of Domain Knowledge}
\label{sec:intro}

The reconstruction of neutrino events in the IceCube detector has traditionally relied on maximum-likelihood based methods~\cite{EnergyReconstruction, AmandaRecos}.
More recently, deep learning architectures such as convolutional neural networks (CNN)~\cite{ConvolutionalNetworks_mod}
have surpassed traditional reconstruction methods in certain areas 
including high-energy cascade reconstruction~\cite{DNNReco}.
Further applications in IceCube, also utilizing recurrent and graph neural networks, are illustrated in Refs.~\cite{NERSC_GNN, IceCubeMunichClassification, ICRC17DeepLearning, IceCubeGNN_martin, FLERCNN_Jessie, FLERCNN_shiqi}.

These deep learning based applications illustrate a paradigm shift from 
explicit to implicit use of available information.
In likelihood based methods, domain knowledge, 
such as translational and rotational invariance, detector characteristics and physics laws,
are implemented directly into the likelihood prescription.
The aforementioned deep learning applications, however, must learn this information implicitly through the training data.
CNNs are able to explicitly utilize translational invariance to a certain degree, 
but they lack the ability to directly include other information.

The strength of deep learning lies in the universality of its methods.
Typical deep learning architectures were developed for a generalized usage in a wide field of applications.
While this enables the application across many different domains, 
it neglects potential benefits from explicit exploitation of domain knowledge.
One such application is the field of image recognition.
Input data in this field consist of an array of pixel values.
Due to the broad range of this domain, the underlying generating process
for these pixel values may not be known.
However, certain information, such as scale, rotational and translational invariance,
may still be shared across applications.
Amongst other reasons, CNNs led to a breakthrough~\cite{ImageNetBreakthrough2} 
in this field by exploiting common domain knowledge such as 
the importance of local pixels and translational invariance.

In contrast to image data, 
the underlying generating processes for data in particle physics experiments are well understood.
These experiments often employ extensive simulations, which implies that the physical processes 
and detector response are known to great detail.
Maximum-likelihood based methods aim to utilize the full extent of this information.
However, due to computational limitations, these methods are often forced to apply simplifications
and approximations.
Standard deep learning architectures perform well for these tasks, but they lack the ability to fully
exploit available domain knowledge.
Similarly to CNNs for image recognition, the explicit utilization of domain knowledge in these
architectures may help to advance the field of event reconstruction in particle physics experiments.
 
In this contribution, a hybrid approach is presented that combines the strengths of deep learning with those of maximum-likelihood based methods.
The presented method utilizes a generative neural network to approximate the detector response for 
a given injected event.
Once trained, the generative model may then be used in a traditional maximum-likelihood setting for event reconstruction.
In contrast to standard deep learning architectures, this approach allows for direct exploitation of available domain knowledge.

\section{Domain Knowledge in IceCube}
\label{sec:icecube}

\begin{figure}
  \begin{center}
    \includegraphics[width=0.49\textwidth, keepaspectratio]{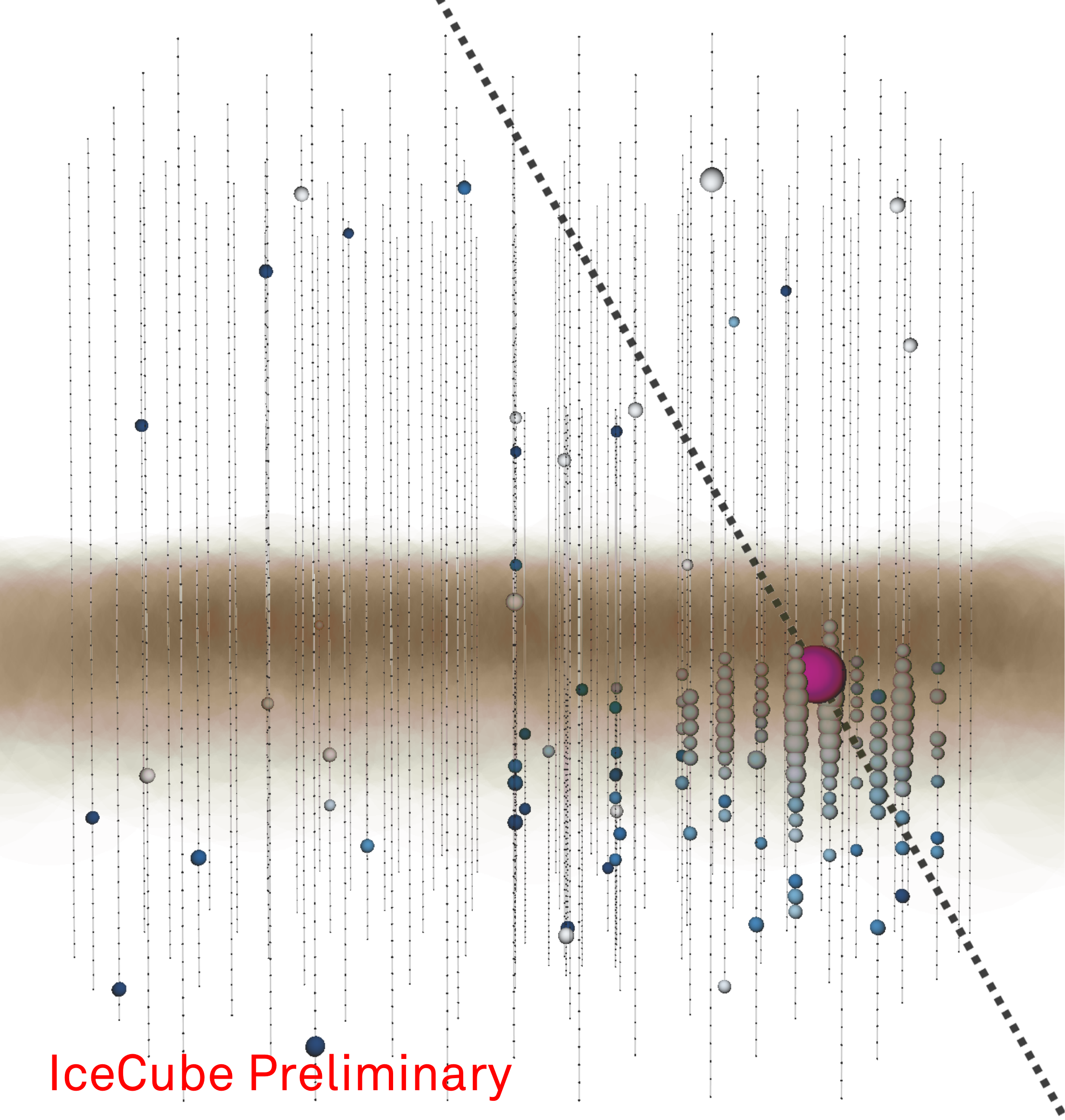}
    \includegraphics[width=0.49\textwidth, keepaspectratio]{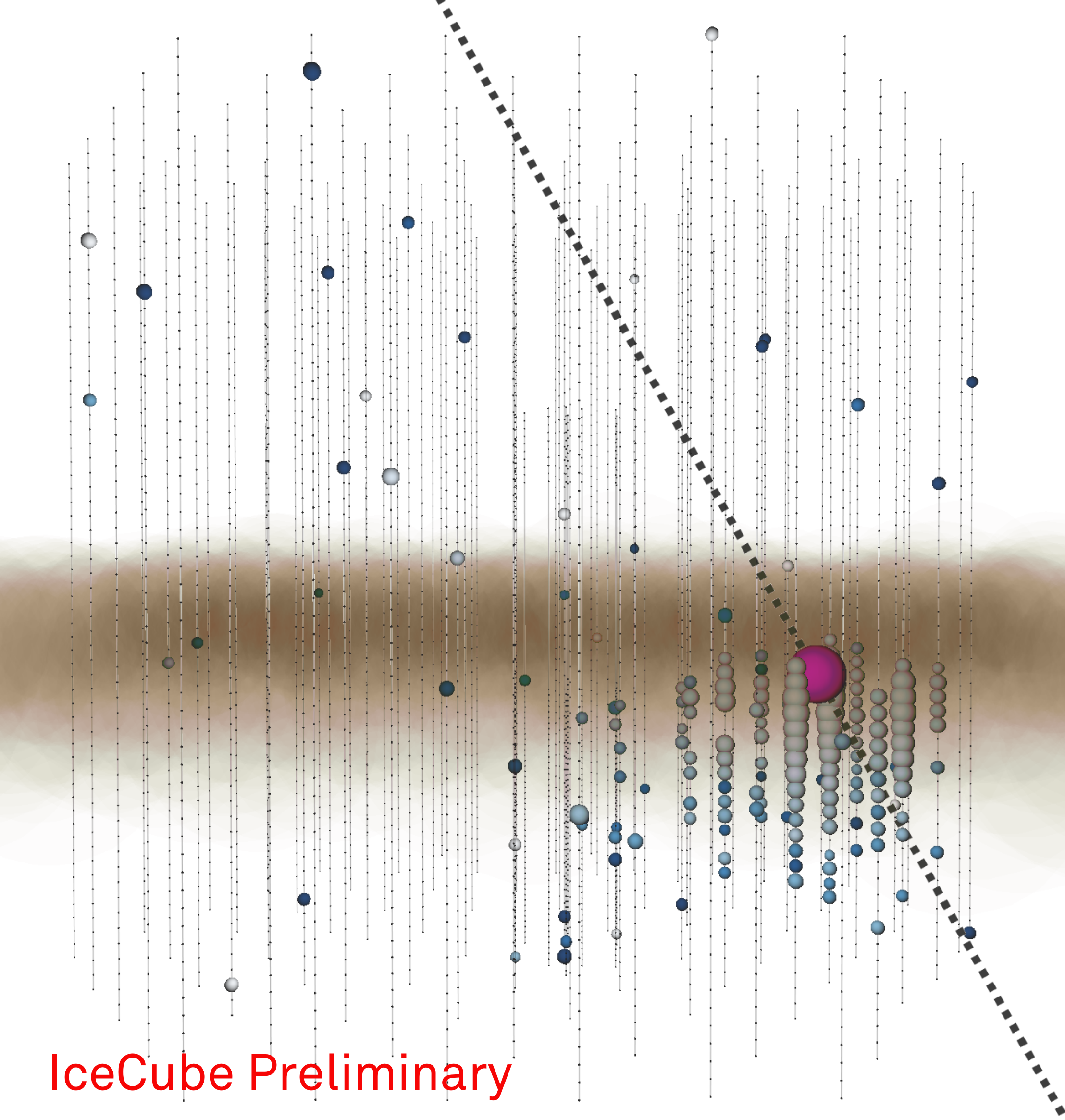}
    \caption{An event view of a cascade interaction just below the dust layer is shown for the Monte Carlo simulation on the left and the generative model on the right. 
    Both events match well within the expected statistical fluctuations.
    The generative model is able to capture the attenuating effect of the dust layer.
    }
    \label{fig:event_views}
  \end{center}
\end{figure}

\begin{figure}
  \begin{minipage}[t]{0.55\textwidth}
    \vspace{0pt}
    \includegraphics[width=\linewidth, keepaspectratio]{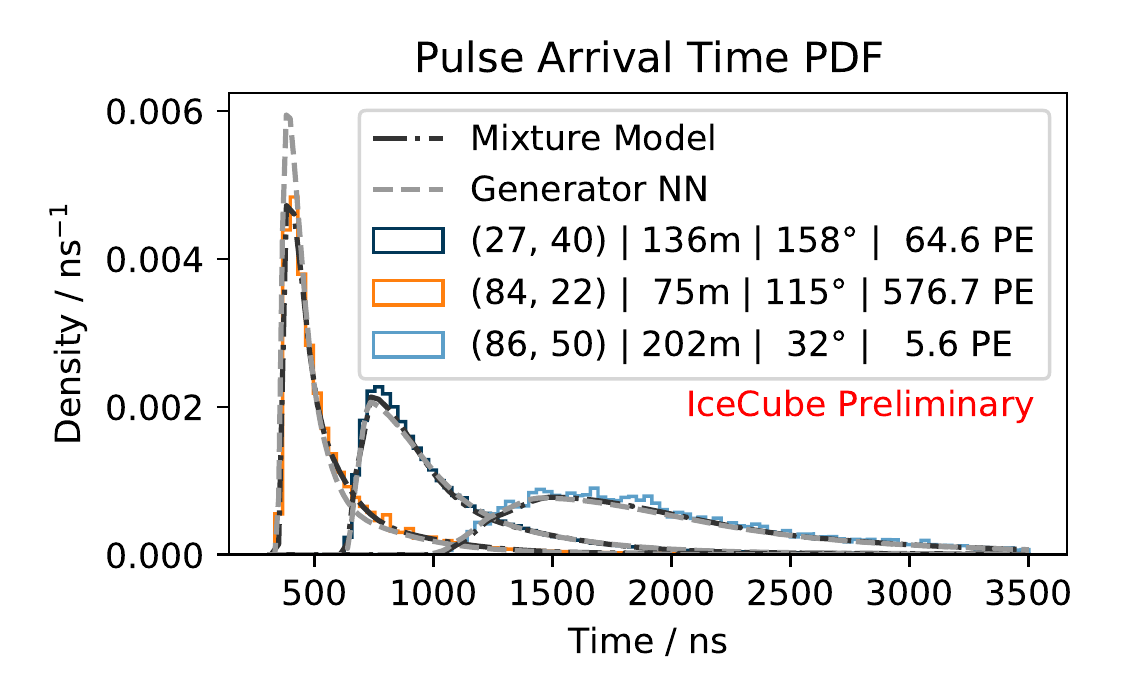}
  \end{minipage}\hfill
  \begin{minipage}[t]{0.45\textwidth}
    \caption{The pulse arrival time PDF at a specific DOM gives insight to the incident angle and the distance of the origin of the arriving photons. 
    The legend indicates the string and DOM number, the distance to the cascade vertex, incident angle, and the average charge collected over 1000 simulations. 
    A mixture model of only three to five asymmetric Gaussians (five shown here) already leads to a good description of the PDF. The generator neural network (NN) is able to model the PDF on unseen data.
    }
    \label{fig:time_pdf}
  \end{minipage}
\end{figure}

IceCube is a cubic-kilometer neutrino detector consisting of 5160 digital optical modules (DOMs) 
installed on 86 strings in the ice at the geographic South Pole
between depths of 1450~$\si{m}$ and 2450~$\si{m}$~\cite{DetectorPaper}. 
Reconstruction of the direction, energy and flavor of the neutrinos relies on the optical detection of Cherenkov radiation emitted by charged particles produced in the interactions of neutrinos in the surrounding ice or the nearby bedrock.
The amount of detected photons scales linearly with deposited energy, and the shape of the pulse arrival
time PDF, as shown in Figure~\ref{fig:time_pdf}, may be used to infer the incident angle and distance of
the particle shower from which the photons originated.
In the context of this contribution, emphasis is put on the reconstruction of cascade events (see Figure~\ref{fig:event_views} for an example event view), which are
induced by interactions of charged current electron neutrinos and neutral current interactions of all neutrino flavors.
%

The geometry of the detector, the translational and rotational invariance of the neutrino interaction, 
the linear scaling between collected charge and deposited energy, and the optical properties of the ice including the dust layer, as depicted in Figure~\ref{fig:event_views}, 
constitute examples of domain knowledge that may be utilized in the event reconstruction task.

The convolutional layers of the CNN based reconstruction method~\cite{DNNReco} assume a regular DOM geometry and translational invariance in the measured pulses, which is only approximately valid.
Irregularities in the hexagonal detector grid, as well as ice properties that break the translational and rotational symmetry in measured data, must be compensated in later stages of the architecture.
Although the CNN performs well, these limitations indicate potential for a better adapted network architecture.

The standard maximum-likelihood based event reconstruction is able to explicitly utilize more domain knowledge than the CNN, but due to computational constraints, it is forced to use simplifications.
The pulse arrival time PDF for a given cascade is obtained from splines fit 
to tabulated Monte Carlo (MC) simulations.
Due to the high dimensionality,
simplifications, such as the approximate azimuthal symmetry, must be used to reduce the
size of the look-up tables.

%
%
%
%
%
%

\section{Combining Maximum-Likelihood and Deep Learning}
\label{sec:event_generator}

%
%
The main limitation of the standard maximum-likelihood method lies in its inability to efficiently model
the pulse arrival time PDF and expected charge at each DOM for a given cascade hypothesis.
The employed look-up table scales poorly with increasing dimensions.
Neural networks, on the other hand, are universal approximators that excel at interpolating high-dimensional data.
A hybrid reconstruction method is defined~\cite{CascadeGenerator} that makes use of this property, by replacing the look-up tables with a generative neural network.
The generative model~$G$
\begin{equation}
  G(\vec{\xi}) = \{\vec{\lambda}, \vec{P}(t)\}
\end{equation}
is trained to map the cascade hypothesis~$\vec{\xi} = (x, y, z, \theta, \Phi, E, t)$ to the expected charge~$\vec{\lambda}$ and pulse arrival time PDF~$\vec{P}(t)$ at each DOM.
The pulse arrival time PDF~$P_d(t)$ at the $d$-th DOM is parameterized by a mixture model
\begin{equation}
  P_d(t) = \sum_j^K w_j \cdot \mathrm{AG}(t | \mu_{(d, j)}, \sigma_{(d, j)}, r_{(d, j)})
\end{equation}
of $K$ asymmetric Gaussians~\cite{AsymmetricGaussian_mod}:
\begin{align}
\mathrm{AG}(x | \mu, \sigma, r) &=  N \cdot \begin{cases}
    \exp{\left(-\frac{(x - \mu)^2}{2\sigma^2}\right)},\qquad x \leq \mu\\
    \exp{\left(-\frac{(x - \mu)^2}{2(\sigma r)^2}\right)},\qquad \mathrm{otherwise}
  \end{cases} \\
   N &= \frac{2}{\sqrt{2 \pi} \cdot \sigma (r + 1)}
\end{align}
where $r$ parameterizes the asymmetry.
The mixture model allows for a good description of the PDF, while keeping the number of free parameters reasonably low as illustrated in Figure~\ref{fig:time_pdf}.

\begin{figure}
  \begin{center}
    \includegraphics[width=1.0\textwidth, keepaspectratio]{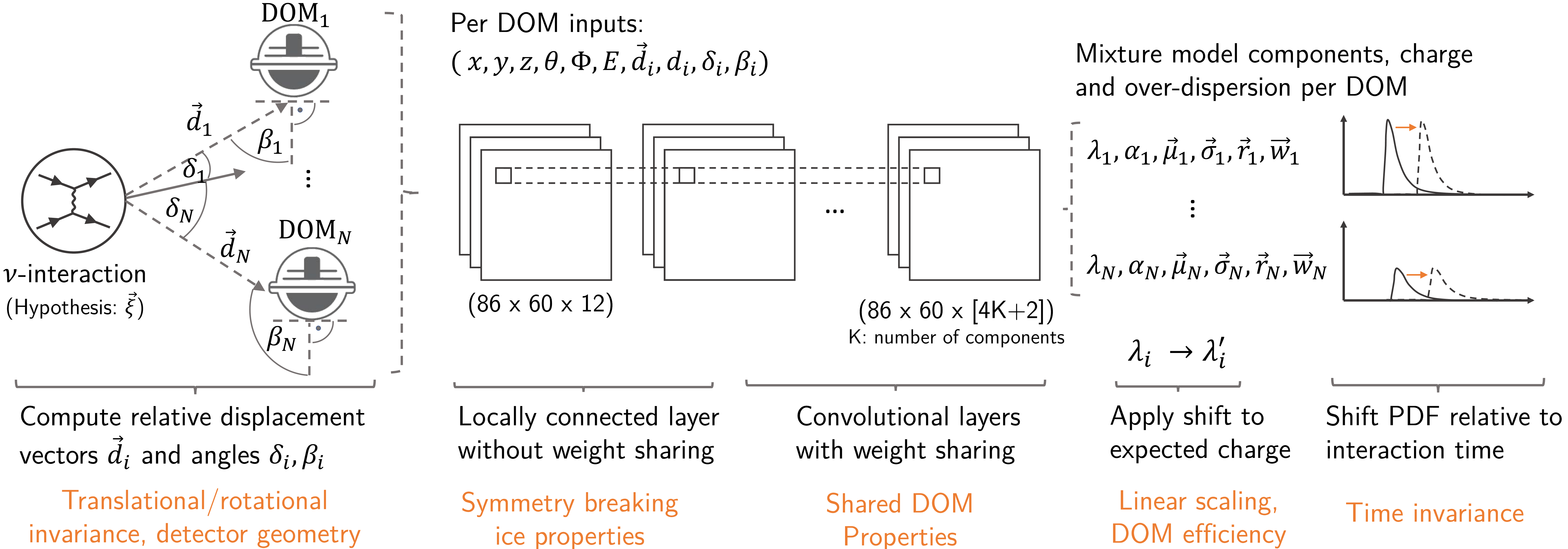}
    \caption{A sketch of the generator neural network architecture is shown. Due to the construction in "forward direction", similar to the Monte Carlo simulation, domain knowledge (examples indicated in orange) can be explictily included into the architecture.}
    \label{fig:architecture}
  \end{center}
\end{figure}

The architecture of the generator NN is therefore setup to output 
the parameters of the mixture model~$\{\vec{\mu}_d, \vec{\sigma}_d, \vec{r}_d, \vec{w}_d\}$ 
and expected charge~$\lambda_d$ for each DOM.
In order to utilize the exact detector geometry and rotational and translational invariance in physics parameter space
of the neutrino interaction, 
relative displacement vectors and angles to each individual DOM are computed and 
provided as input to the NN, as illustrated in Figure~\ref{fig:architecture}.
The neural network performs a series of convolutional layers with $1\times 1$-kernels.
Internally this is implemented in the tensorflow framework~\cite{tensorflow2015-whitepaper} via two dimensional convolutions.
The first layer uses locally connected layers, i.e. it does not apply weight sharing across DOMs.
This allows the NN to model the position and direction dependent, symmetry breaking optical properties of the ice.
Subsequent layers utilize standard convolution operations with weight sharing.
Therefore, after the initial locally connected layer, every DOM is treated equally.
Additional domain knowledge, such as the linear scaling of collected charge to cascade energy or
the differing quantum efficiency~$\epsilon_d$ of the DOMs, is directly incorporated into the architecture,
by scaling the expected charge output:
\begin{equation}
  \lambda_d' = \lambda_d \cdot \frac{E}{\SI{10}{TeV}} \cdot \epsilon_d.
\end{equation}
In general, the architecture may be configured analogously to the MC simulation, 
while computationally expensive parts are replaced by a neural network approximation.
Any domain knowledge that goes into the MC simulation, may therefore also be utilized in the
generator NN.
This is possible in contrast to standard deep learning architectures, 
because the generator NN is defined in the same "forward" direction as the simulation.
Standard deep learning applications, such as the CNN based method, attempt to infer the 
posterior distribution of the quantities of interest from measured data, 
i.e. in "backward" direction compared to the simulation.

Although the focus in this contribution is on the reconstruction of cascades,
this method can be generalized to arbitrary light sources.
In IceCube, any event topology can be built up from a linear superposition of cascades and track segments,
such that only two generative models for these elementary source types are required.
Systematic uncertainties may also be included in the event hypothesis~$\vec{\xi}$ as nuisance parameters.
An alternative method to account for systematic uncertainties, is to marginalize over these during the
training process of the generator NN.
This is accomplished by utilizing a traning dataset that employs the SnowStorm~\cite{SnowStorm} method, 
which samples new systematic parameters from a continuous prior distribution 
for every batch of simulated events.

\begin{figure}
  \begin{minipage}[t]{0.55\textwidth}
    \vspace{0pt}
    \includegraphics[width=\linewidth, keepaspectratio]{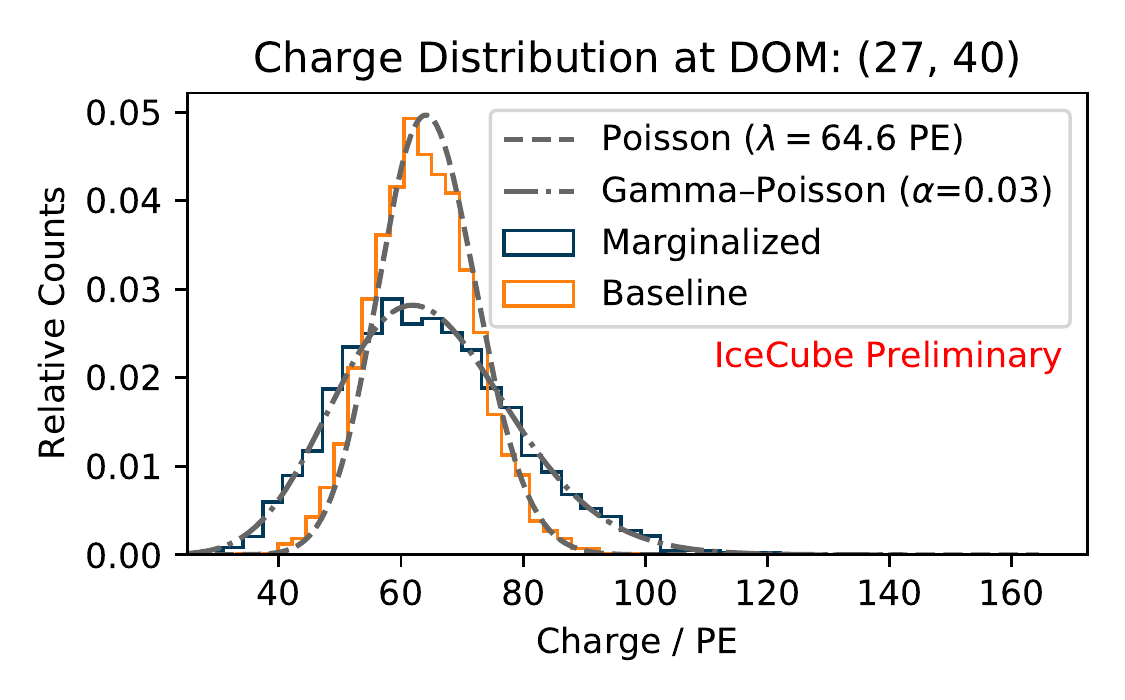}
  \end{minipage}\hfill
  \begin{minipage}[t]{0.4\textwidth}
    \vspace{15pt}
    \caption{The distribution of total charge at a given DOM, obtained from 1000 simulation runs, follows a Poisson distribution for a fixed set of systematic parameters (Baseline). When marginalizing over systematic parameters (Marginalized), the resulting over-dispersion may be modeled via a Gamma-Poisson distribution.}
    \label{fig:charge_distribution}
  \end{minipage}
\end{figure}

For the training procedure, an extended unbinned likelihood over the measured pulses is used.
For the case without systematic parameters or systematic parameters as nuisance parameters, the per-event likelihood is defined as
\begin{equation}
  \mathcal{L}_\mathrm{event}\left( \vec{x}=\{\vec{c}, \vec{t}\} \,|\, \vec{\xi} \right) = 
    \prod_d^D \mathsf{Poisson}\left(\sum_i^{N_d} c_{d,i} \,|\, \lambda_d\left(\vec{\xi}\right)\right) 
    \cdot \prod_i^{N_d}P_d(t_{d,i}\,|\, \vec{\xi})^{c_{d,i}},
    \label{eqn:likelihood1}
\end{equation}
where $D=5160$ is the total number of DOMs, $N_d$ is the number of pulses at the $d$-th DOM, and $c_{d,i}$ and $t_{d,i}$ are the charge and time
of the $i$-th pulse at the $d$-th DOM.
When instead marginalizing over systematics, one must account for the over-dispersion in measured charge.
In this case, the measured charge at a DOM does not follow a Poisson distribution anymore.
A Gamma-Poisson mixture distribution may be used as shown in Figure~\ref{fig:charge_distribution}.
The Gamma-Poisson mixture distribution is a real-valued pendant to the negative binomial distribution
that is capable of modeling the over-dispersion.
The parameterization from Ref.~\cite{SALINAS20201181} is chosen:
\begin{equation}
  \mathsf{GammaPoisson}\left(z | \lambda, \alpha\right) = 
    \frac{\Gamma(z + \frac{1}{\alpha})}{\Gamma(z+1)\Gamma(\frac{1}{\alpha})}
    \left( \frac{1}{1 + \alpha \lambda} \right)^\frac{1}{\alpha} 
    \left( \frac{\alpha\lambda}{1 + \alpha\lambda} \right)^z
\end{equation}
which introduces the shape parameter~$\alpha$ that leads to over-dispersion when $\alpha > 0$.
As a result, the generator NN must also output the shape parameter~$\vec{\alpha}$ for each DOM and the
likelihood is modified to:
\begin{equation}
  \mathcal{L}_\mathrm{event}\left( \vec{x}=\{\vec{c}, \vec{t}\} \,|\, \vec{\xi} \right) = 
    \prod_d^D \mathsf{GammaPoisson}\left(
      \sum_i^{N_d} c_{d,i} \,|\, 
        \lambda_d(\vec{\xi}),
        \alpha_d(\vec{\xi})
    \right) 
    \cdot \prod_i^{N_d}P_d(t_{d,i}\,|\, \vec{\xi})^{c_{d,i}}.
    \label{eqn:likelihood2}
\end{equation}


\subsection*{Model Performance and Applications}


\begin{figure}
  \begin{center}
    \includegraphics[width=0.49\textwidth, keepaspectratio]{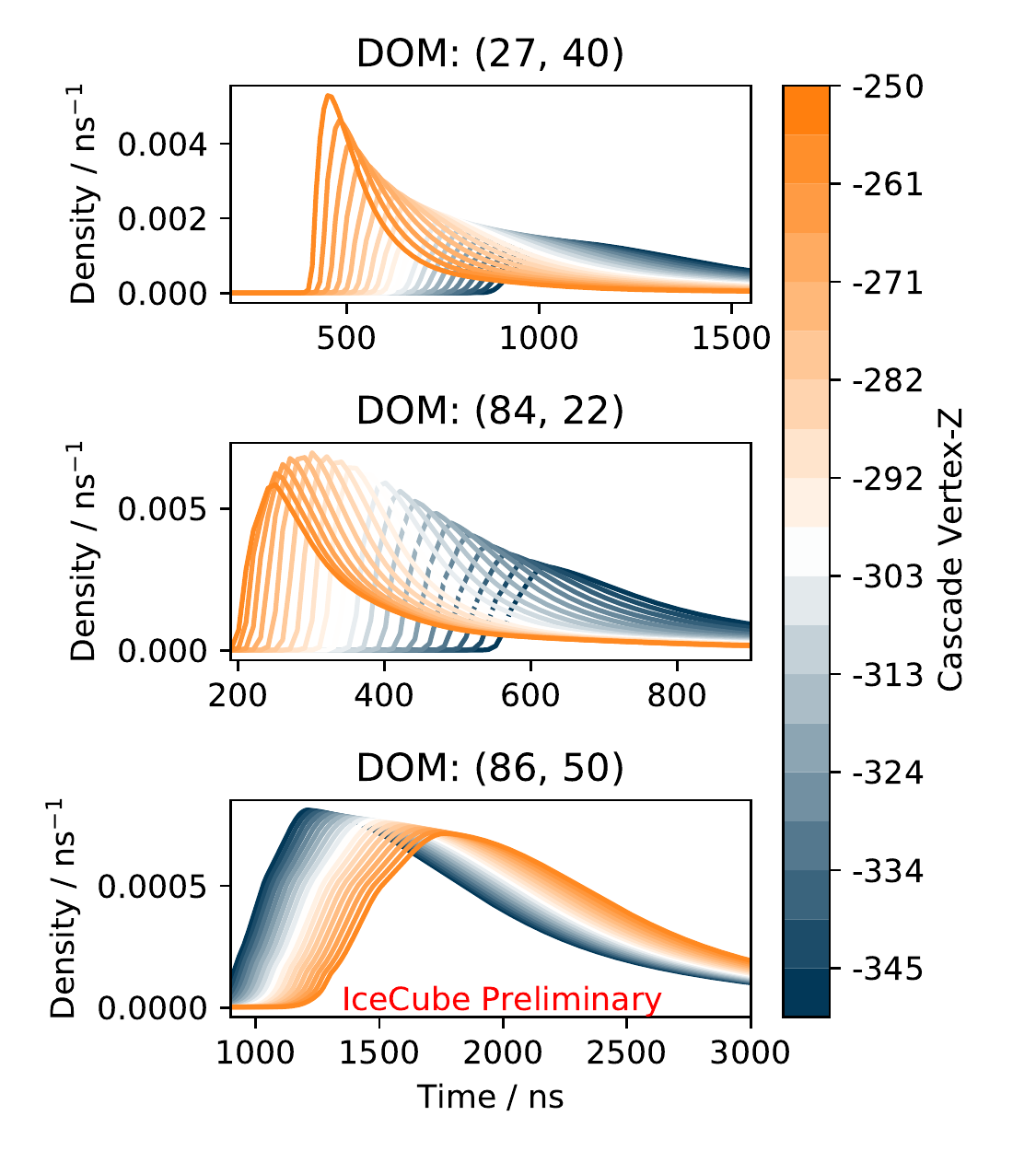}
    \includegraphics[width=0.49\textwidth, keepaspectratio]{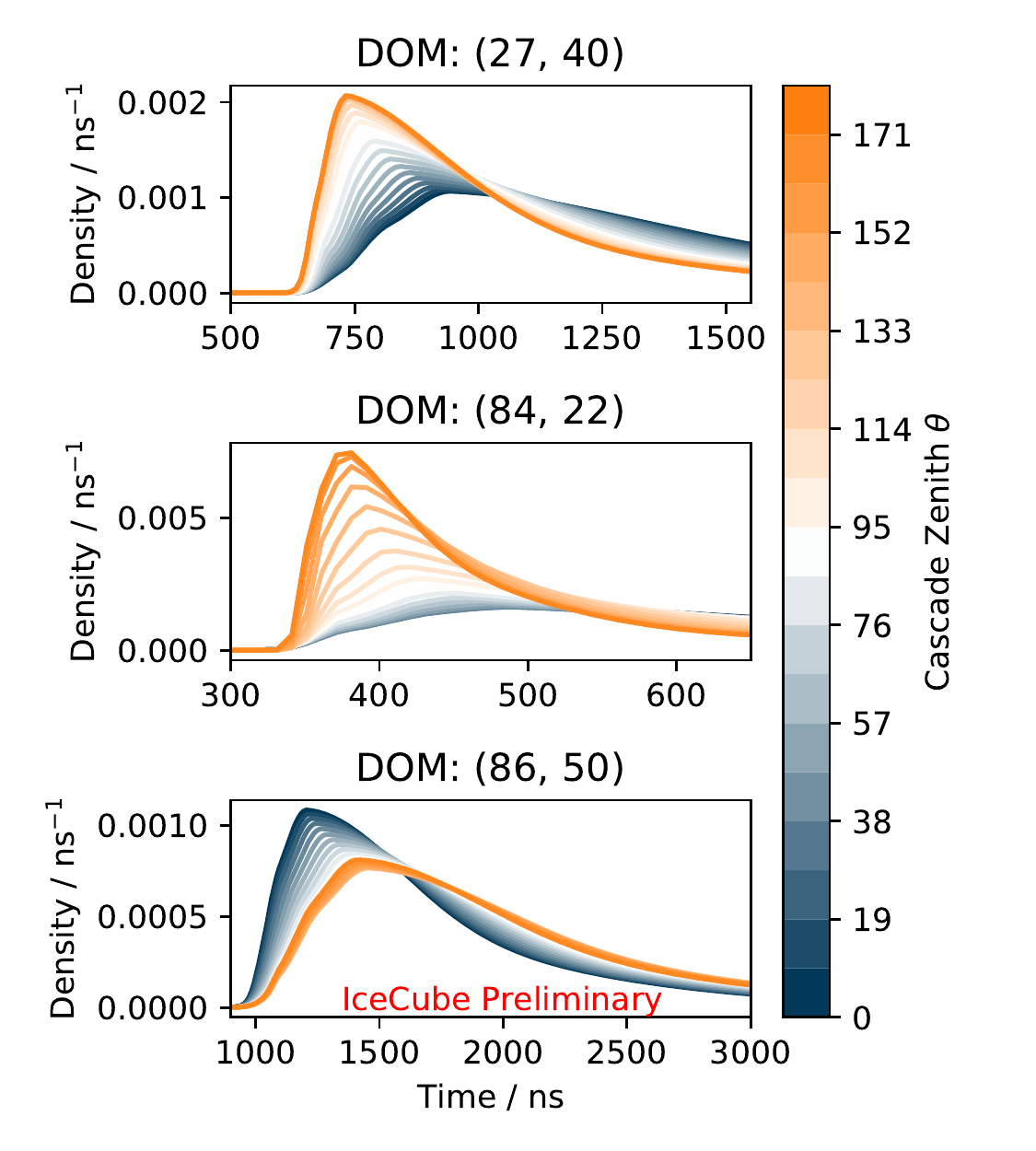}
    \caption{The pulse arrival time PDF, approximated by the generative model, is shown for three different DOMs of the same event. The left panel shows the effect of modifying the $z$-coordinate of the cascade interaction vertex, while the right panel illustrates the change due to the varying zenith angle.}
    \label{fig:scan}
  \end{center}
\end{figure}

\begin{figure}
  \begin{center}
    \includegraphics[width=\textwidth, keepaspectratio]{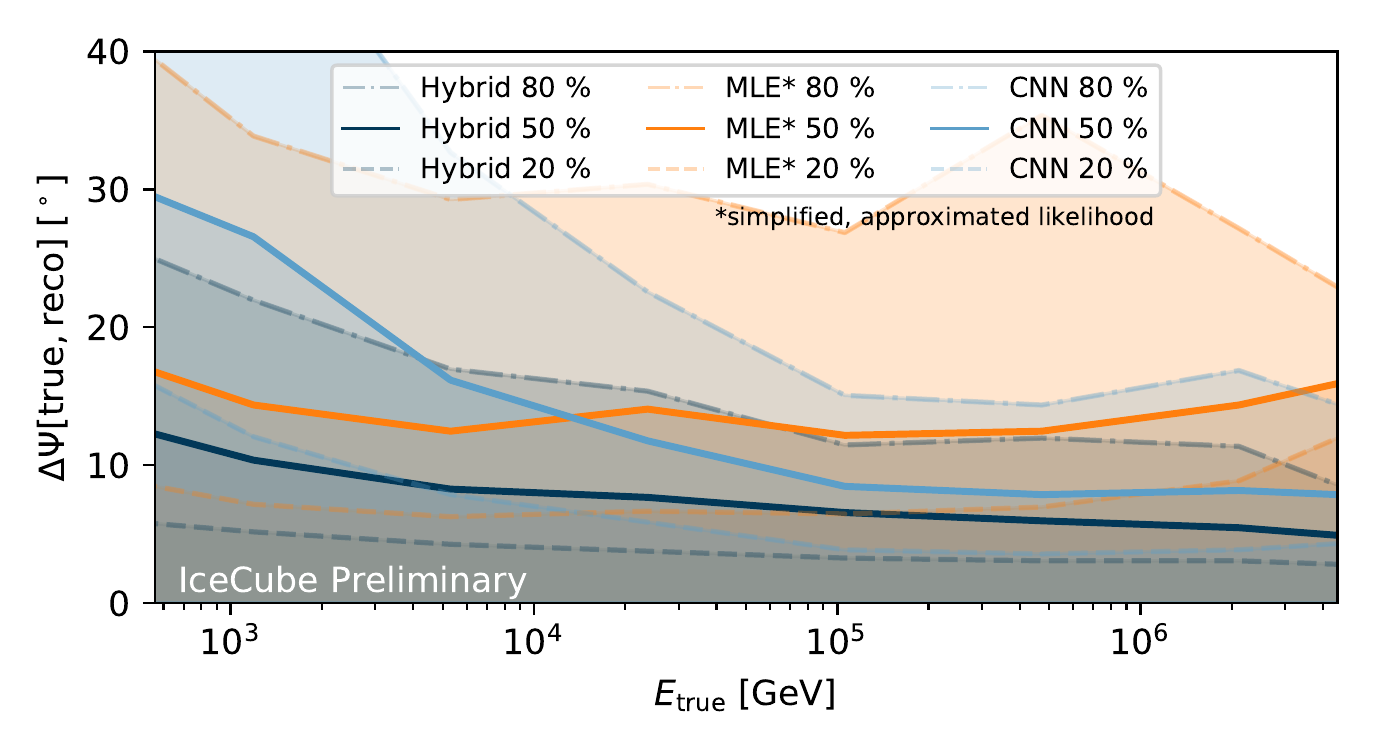}
    \caption{The angular resolution of charged-current NuE interactions for a typical cascade event selection is compared between IceCube's default reconstruction (MLE)~\cite{EnergyReconstruction}, a CNN based method~\cite{DNNReco} and the newly developed hybrid method.
    The hybrid method leads to a significantly improved angular resolution over the whole energy range.
    The plateau towards higher energies is induced by systematic uncertainties.
    }
    \label{fig:resolution}
  \end{center}
\end{figure}

An additional benefit of the generator NN over the standard deep learning architectures lies
in the improved interpretability of the model. 
Individual components of the model may be investigated and cross-checked.
Figure~\ref{fig:time_pdf} demonstrates that the model is capable of correctly modeling the
arrival time PDF on unseen data.
It is also possible to investigate how the PDFs change for individual DOMs 
when modifying the event hypothesis.
In Figure~\ref{fig:scan}, an example is shown in which the cascade zenith values and the
$z$-coordinate of the interaction vertex are shifted.
As expected, the generator NN models a smooth transition from one hypothesis to another.

The trained generative model may be used in a maximum-likelihood setting to reconstruct events via the likelihoods provided in Equations~\eqref{eqn:likelihood1} and~\eqref{eqn:likelihood2}.
The novel hybrid method is able to improve upon the CNN and the standard reconstruction method
over the whole energy range, leading to a significant boost in angular resolution 
(see Figure~\ref{fig:resolution}).
This is possible, because the hybrid method is not subject to simplifications and because it
can benefit from available domain knowledge.
The generative model can also be employed to simulate events as illustrated in the right panel
of Figure~\ref{fig:event_views}.
Other applications such as likelihood scans and Markov-Chain Monte Carlos are also possible.

\section{Conclusions}
\label{sec:conclusions}

A novel hybrid reconstruction method is presented that combines the strengths
of deep learning with those of maximum-likelihood.
This is accomplished by utilizing a generator NN to approximate the high-dimensional
likelihood.
Due to the construction in "forward direction", domain knowledge implemented in the
MC simulation may easily be incorporated in the neural network architecture.
The resulting generative model is a versatile tool that may also be applied in other
applications such as event simulation.

%
%
%
 
\bibliographystyle{ICRC}
\bibliography{references}

\providecommand{\href}[2]{#2}\begingroup\raggedright\begin{thebibliography}{10}

\bibitem{EnergyReconstruction}
{\bfseries IceCube} Collaboration, M.~Aartsen {\em et~al.}
  \href{http://dx.doi.org/10.1088/1748-0221/9/03/P03009}{{\em JINST} {\bfseries
  9} (2014) P03009}.

\bibitem{AmandaRecos}
{\bfseries AMANDA} Collaboration, J.~Ahrens {\em et~al.}
  \href{http://dx.doi.org/10.1016/j.nima.2004.01.065}{{\em Nucl. Instrum. Meth.
  A} {\bfseries 524} (2004) 169--194}.

\bibitem{ConvolutionalNetworks_mod}
Y.~LeCun {\em et~al.} {\em Advances in Neural Information Processing Systems 2}
  (1990) 396--404.

\bibitem{DNNReco}
{\bfseries IceCube} Collaboration, R.~Abbasi {\em et~al.}
  \href{http://dx.doi.org/10.1088/1748-0221/16/07/p07041}{{\em JINST}
  {\bfseries 16} no.~07, (Jul, 2021) P07041}.

\bibitem{NERSC_GNN}
{\bfseries IceCube} Collaboration, N.~Choma {\em et~al.}
  \href{http://arxiv.org/abs/1809.06166}{{\ttfamily arXiv:1809.06166 [cs.LG]}}.

\bibitem{IceCubeMunichClassification}
{\bfseries IceCube} Collaboration, M.~Kronmueller and T.~Glauch
  \href{http://dx.doi.org/10.22323/1.358.0937}{{\em PoS} {\bfseries ICRC2019}
  (2020) 937}.

\bibitem{ICRC17DeepLearning}
{\bfseries IceCube} Collaboration, M.~Huennefeld
  \href{http://dx.doi.org/10.22323/1.301.1057}{{\em PoS} {\bfseries ICRC2017}
  (2018) 1057}.

\bibitem{IceCubeGNN_martin}
{\bfseries IceCube} Collaboration {\em PoS} {\bfseries ICRC2021} (these
  proceedings) 1044.

\bibitem{FLERCNN_Jessie}
{\bfseries IceCube} Collaboration {\em PoS} {\bfseries ICRC2021} (these
  proceedings) 1053.

\bibitem{FLERCNN_shiqi}
{\bfseries IceCube} Collaboration {\em PoS} {\bfseries ICRC2021} (these
  proceedings) 1054.

\bibitem{ImageNetBreakthrough2}
A.~Krizhevsky, I.~Sutskever, and G.~E. Hinton
  \href{http://dx.doi.org/10.1145/3065386}{{\em Commun. ACM} {\bfseries 60}
  no.~6, (May, 2017) 84--90}.

\bibitem{DetectorPaper}
{\bfseries IceCube} Collaboration, M.~Aartsen {\em et~al.}
  \href{http://dx.doi.org/10.1088/1748-0221/12/03/P03012}{{\em JINST}
  {\bfseries 12} no.~03, (2017) P03012}.

\bibitem{CascadeGenerator}
{\bfseries IceCube} Collaboration, M.~Huennefeld
  \href{http://dx.doi.org/10.1051/epjconf/201920705005}{{\em EPJ Web Conf.}
  {\bfseries 207} (2019) 05005}.

\bibitem{AsymmetricGaussian_mod}
T.~Kato, S.~Omachi, and H.~Aso
  \href{http://dx.doi.org/10.1007/3-540-70659-3_42}{{\em Lecture Notes in
  Computer Science} (2002) 405--413}.

\bibitem{tensorflow2015-whitepaper}
M.~Abadi {\em et~al.} \href{http://arxiv.org/abs/1603.04467}{{\ttfamily
  arXiv:1603.04467}}.

\bibitem{SnowStorm}
{\bfseries IceCube} Collaboration, M.~Aartsen {\em et~al.}
  \href{http://dx.doi.org/10.1088/1475-7516/2019/10/048}{{\em JCAP} {\bfseries
  10} (2019) 048}.

\bibitem{SALINAS20201181}
D.~Salinas {\em et~al.}
  \href{http://dx.doi.org/https://doi.org/10.1016/j.ijforecast.2019.07.001}{{\em
  International Journal of Forecasting} {\bfseries 36} no.~3, (2020)
  1181--1191}.

\end{thebibliography}\endgroup

%

\clearpage
\section*{Full Author List: IceCube Collaboration}




\scriptsize
\noindent
R. Abbasi$^{17}$,
M. Ackermann$^{59}$,
J. Adams$^{18}$,
J. A. Aguilar$^{12}$,
M. Ahlers$^{22}$,
M. Ahrens$^{50}$,
C. Alispach$^{28}$,
A. A. Alves Jr.$^{31}$,
N. M. Amin$^{42}$,
R. An$^{14}$
% ...
\\

\noindent
% ...
$^{12}$ Universit{\'e} Libre de Bruxelles, Science Faculty CP230, B-1050 Brussels, Belgium \\
% ...
$^{14}$ Department of Physics and Laboratory for Particle Physics and Cosmology, Harvard University, Cambridge, MA 02138, USA \\
% ...
$^{17}$ Department of Physics, Loyola University Chicago, Chicago, IL 60660, USA \\
$^{18}$ Dept. of Physics and Astronomy, University of Canterbury, Private Bag 4800, Christchurch, New Zealand \\
% ...
$^{22}$ Niels Bohr Institute, University of Copenhagen, DK-2100 Copenhagen, Denmark \\
% ...
$^{28}$ D{\'e}partement de physique nucl{\'e}aire et corpusculaire, Universit{\'e} de Gen{\`e}ve, CH-1211 Gen{\`e}ve, Switzerland \\
% ...
$^{31}$ Karlsruhe Institute of Technology, Institute for Astroparticle Physics, D-76021 Karlsruhe, Germany  \\
% ...
$^{42}$ Bartol Research Institute and Dept. of Physics and Astronomy, University of Delaware, Newark, DE 19716, USA \\
% ...
$^{50}$ Oskar Klein Centre and Dept. of Physics, Stockholm University, SE-10691 Stockholm, Sweden \\
% ...
$^{59}$ DESY, D-15738 Zeuthen, Germany \\
% ...

\subsection*{Acknowledgements}

\noindent
USA {\textendash} U.S. National Science Foundation-Office of Polar Programs,
U.S. National Science Foundation-Physics Division,
U.S. National Science Foundation-EPSCoR,
Wisconsin Alumni Research Foundation,
Center for High Throughput Computing (CHTC) at the University of Wisconsin{\textendash}Madison,
Open Science Grid (OSG),
Extreme Science and Engineering Discovery Environment (XSEDE),
Frontera computing project at the Texas Advanced Computing Center,
U.S. Department of Energy-National Energy Research Scientific Computing Center,
Particle astrophysics research computing center at the University of Maryland,
Institute for Cyber-Enabled Research at Michigan State University,
and Astroparticle physics computational facility at Marquette University;
Belgium {\textendash} Funds for Scientific Research (FRS-FNRS and FWO),
FWO Odysseus and Big Science programmes,
and Belgian Federal Science Policy Office (Belspo);
Germany {\textendash} Bundesministerium f{\"u}r Bildung und Forschung (BMBF),
Deutsche Forschungsgemeinschaft (DFG),
Helmholtz Alliance for Astroparticle Physics (HAP),
Initiative and Networking Fund of the Helmholtz Association,
Deutsches Elektronen Synchrotron (DESY),
and High Performance Computing cluster of the RWTH Aachen;
Sweden {\textendash} Swedish Research Council,
Swedish Polar Research Secretariat,
Swedish National Infrastructure for Computing (SNIC),
and Knut and Alice Wallenberg Foundation;
Australia {\textendash} Australian Research Council;
Canada {\textendash} Natural Sciences and Engineering Research Council of Canada,
Calcul Qu{\'e}bec, Compute Ontario, Canada Foundation for Innovation, WestGrid, and Compute Canada;
Denmark {\textendash} Villum Fonden and Carlsberg Foundation;
New Zealand {\textendash} Marsden Fund;
Japan {\textendash} Japan Society for Promotion of Science (JSPS)
and Institute for Global Prominent Research (IGPR) of Chiba University;
Korea {\textendash} National Research Foundation of Korea (NRF);
Switzerland {\textendash} Swiss National Science Foundation (SNSF);
United Kingdom {\textendash} Department of Physics, University of Oxford.

\end{document}